\documentclass[sn-mathphys,iicol,Numbered]{sn-jnl}

\usepackage{graphicx}%
\usepackage{multirow}%
\usepackage{amsmath,amssymb,amsfonts}%
\usepackage{amsthm}%
\usepackage{mathrsfs}%
\usepackage[title]{appendix}%
\usepackage{xcolor}%
\usepackage{textcomp}%
\usepackage{manyfoot}%
\usepackage{booktabs}%
\usepackage{algorithm}%
\usepackage{algorithmicx}%
\usepackage{algpseudocode}%
\usepackage{listings}%
\usepackage{mathtools, cuted}
\usepackage{balance}

\begin{document}

\title[]{Physics and data driven model for prediction of residual stresses in machining}

\author*[1,2]{\fnm{Rachit} \sur{Dhar}}\email{ms1200760@iitd.ac.in}

\author[1]{\fnm{Ankur} \sur{Krishna}}\email{ankur.krishna@tcs.com}

\author[1]{\fnm{Bilal} \sur{Muhammed}}\email{bilal.muhammed@tcs.com}

\affil[1]{\orgdiv{TCS Research}, \orgname{Tata Consultancy Services}, \orgaddress{\city{Pune}, \state{Maharashtra}, \country{India}}}

\affil[2]{\orgdiv{Department of Materials Science and Engineering}, \orgname{Indian Institute of Technology Delhi}, \orgaddress{\postcode{110016}, \city{New Delhi}, \country{India}}}

\abstract{Predicting residual stresses has always been a topic of significance due to its implications in the development of enhanced materials and better processing conditions. In this work, an analytical model for prediction of residual stresses is developed for orthogonal machining. It consists of three component models for force, temperature and stress computation. The Oxley force model and Waldorf's slip-line model are employed for obtaining cutting force, thrust force, and temperatures at the shear zone and tool-chip interface for the given parameters. The Komanduri-Hou two heat source model is used for obtaining the temperature distribution in the workpiece. The effect of coolant with differing mass flow rates has also been incorporated. The residual stresses are obtained by combining the mechanical and thermal components, followed by the loading and relaxation of the stresses. Optimal values for unknown parameters are predicted by leveraging a cost function. The residual stress distributions obtained give a tensile region near the surface for Inconel 718, and a compressive region for Ti6Al4V, which are in line with experimental results found in literature.}


\maketitle

\section{Introduction}

Understanding and predicting residual stresses has been a significant focus of research because controlling their formation can lead to more durable and safer materials. Therefore, extensive research has been conducted on modeling residual stress distributions in materials and studying the effects of different materials and process parameters. Residual stresses in orthogonal machining arise from two main factors: (a) local plastic deformation during cutting, which persists after the cutting forces are removed, and (b) thermal stresses due to heat generation at the tool-chip interface and in the shear zone.

Initial work on residual stress modeling relied on empirical models, but their applicability is limited to specific conditions and cannot be extended to other situations \cite{wanreview}. Analytical models have been developed to address this limitation by providing a general framework capable of handling various parameters based on the underlying mechanisms of residual stress generation \cite{wanreview}. Merwin and Johnson \cite{merwinjohnson} proposed an analytical model based on classical Hertz theory for calculating stress distribution, which was later refined by Johnson \cite{johnson}. McDowell \cite{mcdowell} used cycloidal distributions to describe the functions $p(s)$ and $q(s)$ in the Johnson stress model. Su and Liang \cite{liangsu} utilized Oxley's \cite{oxley} cutting force model and Waldorf's \cite{waldorf} plowing model to predict residual stresses analytically. Wan et al. \cite{wan} modified the plowing force model by incorporating the Johnson-Cook model \cite{johnsoncook} for flow shear stress calculation, which became widely adopted. Initial experimental work on temperature distribution was conducted by Matsumoto et al. \cite{matsumoto}, and a temperature model was developed by Jacobus et al. \cite{jacobus}. Sekhon and Chenot \cite{sekhon} derived an expression for $\gamma$, the friction heat partition coefficient. Ulutan et al. \cite{ulutan} employed a finite-difference model for temperature distribution. Komanduri and Hou \cite{komanduri1, komanduri2, komanduri3} proposed a two heat source model. Pan et al. \cite{pan} suggested a modified Johnson-Cook model for analytical residual stress modeling. Drucker and Palgen \cite{drucker} provided an expression for the plastic modulus. Sehitoglu and Jiang \cite{sehitoglu} introduced a non-linear kinematic hardening model for elasto-plastic loading, later modified by McDowell \cite{mcdowell}. Merwin and Johnson \cite{merwinjohnson} provided boundary conditions to eliminate certain components of the residual stress tensor based on the assumption of plane deformation. Finite element methods (FEMs) represent a newer set of techniques for simulating tool-workpiece interactions without requiring deep mechanistic analysis. However, FEMs often suffer from low efficiency and high computational demands \cite{wanreview}. Therefore, analytical models remain the most suitable choice for meeting industrial needs.

This paper presents an analytical model for predicting residual stresses in orthogonal machining, comprising three component analytical models for force, temperature, and stress. To enhance accuracy, optimal values of unknown parameters ($C_o$, $\delta$ in the force model, and the plastic modulus function $h$ in stress loading and relaxation) are determined within a specific range based on a cost function to minimize errors in force and stress prediction. The Oxley cutting force model is used to predict cutting and thrust forces, along with Waldorf's model for plowing forces, as reported by Su \cite{su}, Zhou and Ren \cite{zhou-in}. Temperatures in the shear zone and tool-chip interface are determined using Boothroyd's \cite{boothroyd} cutting temperature model. The Johnson-Cook model is employed to model flow stresses. The temperature distribution across the workpiece is obtained using the Komanduri-Hou model, incorporating the additional effect of a coolant based on the work of Singh and Sharma \cite{singh}. This temperature profile is then used to model mechanical \cite{johnson} and thermal stresses \cite{liang1, liang2}, which are combined and incrementally relaxed following the methods outlined by Shan et al. \cite{shan}, Liang et al. \cite{liang1, liang2}, McDowell \cite{mcdowell}, and Merwin and Johnson \cite{merwinjohnson}, ultimately yielding the residual stresses. The results of forces, temperature, and residual stress are plotted and discussed for Inconel 718 and Ti6Al4V machining data from the literature.

The paper proceeds by discussing various models used, starting with the force model in Section 2 for calculating cutting and thrust forces, followed by the temperature distribution model in Section 3, and concluding with the residual stress model in Section 4. The parameters used for validating the model are presented in Section 5, while Section 6 discusses the results and their implications when applying the model to machining data for two materials - Inconel 718 and Ti6Al4V. Finally, the conclusions drawn from the study are discussed in Section 7.

\begin{figure}
	\includegraphics[width=0.5\textwidth]{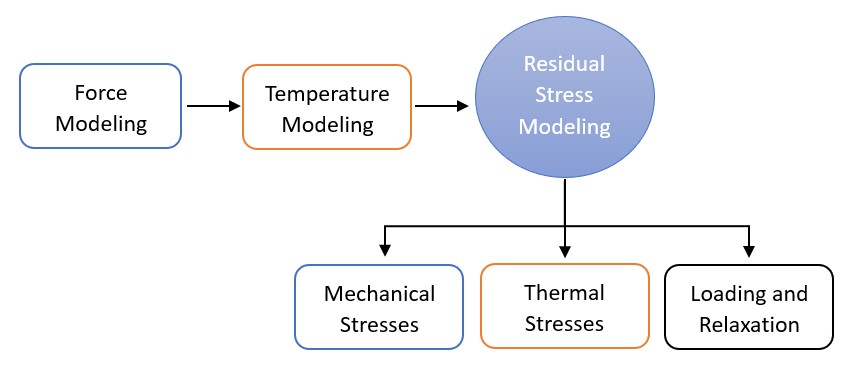}
	\caption{Overall workflow of the paper}
	\label{fig:1}
\end{figure}

\section{Force modeling for orthogonal cutting}

To determine the mechanical stresses and temperature changes within the workpiece during orthogonal machining, it is necessary to ascertain the cutting and thrust forces. These forces are computed using Oxley's model, which in turn relies on flow stresses derived from the Johnson-Cook model. The following section provides a detailed description of these models.

\subsection{Oxley cutting force model}

Oxley \cite{oxley} introduced a model to compute the forces encountered in orthogonal machining. Su \cite{su} presented an algorithm for utilizing Oxley's model to calculate forces. In accordance with their research, the resultant force $R$, as well as the shear cutting and thrust forces ($F_c^{shear}$ and $F_t^{shear}$), are expressed by the following equations:

\begin{equation}
	R = \dfrac{k_{AB} t w}{\sin\phi \cos\theta}
\end{equation}

\begin{equation}
	F_c^{shear} = R \cos(\lambda - \alpha)
\end{equation}

\begin{equation}
	F_t^{shear} = R \sin(\lambda - \alpha)
\end{equation}

where $k_{AB}$ is the flow stress in the shear zone, $t$ is the uncut chip thickness (i.e., depth of cut), and $w$ is the width of cut. $\phi$, $\alpha,$ and $\lambda$ represent the shear angle, tool rake angle, and friction angle respectively. The shear speed $v_s$ and chip speed $v_c$ are obtained using the cutting speed $v$ as:

\begin{equation}
	v_s = \dfrac{v \cos\alpha}{\cos(\phi - \alpha)}
\end{equation}

\begin{equation}
	v_c = \dfrac{v \sin\phi}{\cos(\phi - \alpha)}
\end{equation}

The strain ($\varepsilon_{AB}$) and strain rate ($\dot{\varepsilon}_{AB}$) in the shear zone are:

\begin{equation}
	\varepsilon_{AB} = \dfrac{\lambda^{'} \cos\alpha}{\sqrt{3} \sin\phi \cos(\phi - \alpha)}
\end{equation}

\begin{equation}
	\dot{\varepsilon}_{AB} = \dfrac{C_o v_s}{\sqrt{3} (\dfrac{t}{\sin \phi})}
\end{equation}

where $\lambda^{'}$ is given by $\cos\phi \cos(\phi - \alpha) / \cos\alpha$. The angles $\theta$ and $\lambda$ are obtained from:

\begin{equation}
	\theta = \arctan(1 + 2(\dfrac{\pi}{4} - \phi) - C_o n_{eq})
\end{equation}

\begin{equation}
	\lambda = \theta + \alpha - \phi
\end{equation}

Here, $n_{eq}$ is given by the following expression:

\begin{equation}
	n_{eq} = \dfrac{n B \varepsilon_{AB}^n}{A + B \varepsilon_{AB}^n}
\end{equation}

Also, the chip thickness $t_c$ is given by $t \cos(\phi - \alpha) / \sin\phi$. The plowing forces $P_c$ and $P_t$ are modeled according to Waldorf's slip-line model \cite{waldorf}, the equations for which are given as follows:

\begin{equation}
	\begin{split}
		P_c &= k_{AB} w \bigl[(1 + 2\theta + 2\lambda + \sin(2\eta)) \sin(\phi - \gamma + \eta) \\
		&+ \cos(2\eta)\cos(\phi - \gamma + \eta)\bigr]CB
	\end{split}	
\end{equation}

\begin{equation}
	\begin{split}
		P_t &= k_{AB} w \bigl[(1 + 2\theta + 2\lambda + \sin(2\eta)) \cos(\phi - \gamma + \eta) \\
		&- \cos(2\eta)\sin(\phi - \gamma + \eta)\bigr]CB
	\end{split}
\end{equation}

where $CB = R_{plow} / \sin(\eta_{plow})$. The angles $\eta_{plow}, \gamma_{plow}$, and $\theta_{plow}$ are calculated based on the work developed by Zhou and Ren \cite{zhou-in}. In this paper, the equation for calculating the sector radius $R_{plow}$ has been solved analytically, by taking the maximum of the two quadratic roots, instead of using the iterative approach proposed by Zhou and Ren \cite{zhou-in}. The total cutting force $F_c$ and total thrust force $F_t$ are obtained by:

\begin{equation}
	\begin{split}
		F_c &= F_c^{shear} + P_c \\
		F_t &= F_t^{shear} + P_t
	\end{split}
\end{equation}

\subsection{Boothroyd's cutting temperature model}

Boothroyd \cite{boothroyd} described a model for calculating the average temperatures at the shear zone ($T_{AB}$) and the tool-chip interface ($T_{int}$), which is also proposed by Oxley \cite{oxley}:

\begin{equation}
	T_{AB} = T_0 + \eta \Delta T_{AB}
\end{equation}

\begin{equation}
	T_{int} = T_0 + \Delta T_{AB} + \psi \Delta T_M
\end{equation}

\begin{equation}
	\Delta T_{AB} = \dfrac{1-\beta}{\rho Stw}\dfrac{F_s \cos\alpha}{\cos(\phi - \alpha)}
\end{equation}

where $\Delta T_M$ and $\beta$ are calculated using the work done by Oxley \cite{oxley}. The value of the shear force $F_s$ has been computed using an iterative algorithm given by Zhou and Ren \cite{zhou-in}.

\subsection{Johnson-Cook model}

The flow stresses in the shear zone $k_{AB}$, and in the tool-chip interface $k_{int}$ are obtained using the Johnson-Cook model \cite{johnsoncook}, the equations for which have been shown.

\begin{equation}
	\resizebox{\hsize}{!}{$k_{AB} = \dfrac{1}{\sqrt{3}} (A + B\varepsilon_{AB}^n)\left(1 + C\ln{\dfrac{\dot{\varepsilon}_{AB}}{\dot{\varepsilon}_0}}\right)\left(1 - \left( \dfrac{T_{AB} - T_0}{T_m - T_0} \right)^m\right)$}
\end{equation}

\begin{equation}
	\resizebox{\hsize}{!}{$k_{int} = \dfrac{1}{\sqrt{3}} (A + B\varepsilon_{int}^n)\left(1 + C\ln{\dfrac{\dot{\varepsilon}_{int}}{\dot{\varepsilon}_0}}\right)\left(1 - \left( \dfrac{T_{int} - T_0}{T_m - T_0} \right)^m\right)$}
\end{equation}

The average strain $\varepsilon_{int}$ and strain rate $\dot{\varepsilon}_{int}$ in the chip are approximated by the following expressions:

\begin{equation}
	\varepsilon_{int} = 2\varepsilon_{AB} + \dfrac{h}{\sqrt{3}\delta t_c}
\end{equation}

\begin{equation}
	\dot{\varepsilon}_{int} = \dfrac{v_c}{\sqrt{3}\delta t_c}
\end{equation}

Here, the thickness of the primary shear zone $h$ is given by:

\begin{equation}
	h = \dfrac{t \sin\theta}{\cos\lambda \sin\theta}\left(1 + \dfrac{C_o n_{eq}}{3\tan\theta}\right)
\end{equation}

The shear stress along tool-chip interface is given by $\tau_{int} = F / (hw)$, and the shear angle $\phi$ is chosen such that we obtain $\tau_{int} = k_{int}$.

\section{Temperature modeling}

Apart from the mechanical influences of deformation, stress formation in the material is also influenced by temperature changes. This is determined by integrating the heating effect within the workpiece from the shear zone and tool-chip interface, followed by the cooling effect due to the presence of a coolant.

\subsection{The temperature model}

Komanduri and Hou \cite{komanduri1, komanduri2, komanduri3} developed a two heat source model. Based on this model, the temperature change due to shearing and rubbing interactions are modeled. The temperature change due to coolant is obtained using the work done by Singh and Sharma \cite{singh}. The three equations for $\Delta T_{shear}$, $\Delta T_{rubbing}$ and $\Delta T_{cool}$ are given.

\begin{equation}
	\resizebox{\hsize}{!}{$
		\begin{split}
			\Delta T_{shear}(X, Z) &= \dfrac{q_{shear}}{2\pi k} \int_{0}^{L}e^{-\dfrac{(X-l_i sin\varphi)v}{2a}} \bigl(K_0\bigl[\dfrac{v}{2a}\sqrt{(X-l_i sin\varphi)^2 + (Z-l_i cos\varphi)^2}\bigr]\\
			&+ K_0\bigl[\dfrac{v}{2a}\sqrt{(X-l_i sin\varphi)^2 + (Z+l_i cos\varphi)^2}\bigr]\bigr) dl_i
		\end{split}
	$}
\end{equation}

\begin{equation}
	\resizebox{\hsize}{!}{$
		\begin{split}
			\Delta T_{rubbing}(X, Z) = \dfrac{q_{rubbing}}{\pi k} \int_{0}^{\text{VB}}\gamma e^{-\dfrac{(X-x)v}{2a}} K_0\bigl[\dfrac{v}{2a}\sqrt{(X-x)^2 + Z^2}\bigr] dx
		\end{split}
	$}
\end{equation}

\begin{equation}
	\resizebox{\hsize}{!}{$
		\begin{split}
			\Delta T_{cool}(X, Z) = -\dfrac{q_{cool}}{\pi k} \int_{0}^{l} e^{-\dfrac{(X-x)v}{2a}} K_0\bigl[\dfrac{v}{2a}\sqrt{(X-x)^2 + Z^2}\bigr] dx
		\end{split}
	$}
\end{equation}

Here, $\varphi = \phi - \pi/2$, $L = t / \sin\phi$, $\text{VB}$ is the flank wear length, $a$ is the thermal diffusivity of the workpiece, $k$ is the thermal conductivity of the workpiece, and $K_0$ is the modified Bessel function of the second kind of order zero.

The values for heat sources $q_{shear}$ and $q_{rubbing}$ are given by:

\begin{equation}
	q_{shear} = \dfrac{(F_c \cos\phi - F_t \sin\phi)(v \cos\alpha / \cos(\phi - \alpha))}{tw \csc\phi}
\end{equation}

\begin{equation}
	q_{rubbing} = \dfrac{P_c v}{w(\text{VB})}
\end{equation}

\subsection{The effect of coolant}

For a coolant with heat transfer coefficient $h$, the rate of heat loss through convection per unit area is given by $q_{cool} = h(T - T_0)$, as proposed by Su \cite{su}. For a coolant flowing with a mass flow rate per unit area of $\rho v$, the heat transfer coefficient $h$ is obtained using the following expressions given by Singh and Sharma \cite{singh}:

\begin{equation}
	1.86 {Re}^{1/3} {Pr}^{1/3} = \dfrac{hd}{k}
\end{equation}

\begin{equation}
	0.023 {Re}^{0.8} {Pr}^{0.4} = \dfrac{hd}{k}
\end{equation}

where, $Re = \rho v d / \mu$ is the Reynolds number, $Pr$ is the Prandtl number, and the two equations above are for the cases of laminar and turbulent flow respectively. The overall temperature change is then given by $\Delta T (X, Z) = \Delta T_{shear}(X, Z) + \Delta T_{rubbing}(X, Z) + \Delta T_{cool}(X, Z)$.

\section{Residual stress modeling}

The mechanical and thermal stresses that develop within the material are now calculated. These stresses are combined under either elastic or elasto-plastic loading conditions to derive the components of total stress. Subsequently, they are relaxed to ascertain the residual stress value. The previously obtained temperature distribution is employed in computing thermal stresses.

\subsection{Johnson model for mechanical Stress}

Johnson \cite{johnson} provided a model for obtaining the components of mechanical stresses. These equations for $\sigma_{xx}^m$, $\sigma_{zz}^m$ and $\tau_{xz}^m$ are:

\begin{equation}
	\resizebox{\hsize}{!}{$\sigma_{xx}^m = -\dfrac{2z}{\pi}\int_{-b}^{a}\dfrac{p(s)(x-s)^2}{[(x-s)^2 + z^2]^2}ds -\dfrac{2}{\pi}\int_{-b}^{a}\dfrac{q(s)(x-s)^3}{[(x-s)^2 + z^2]^2}ds$}
\end{equation}

\begin{equation}
	\resizebox{\hsize}{!}{$\sigma_{zz}^m = -\dfrac{2z^3}{\pi}\int_{-b}^{a}\dfrac{p(s)}{[(x-s)^2 + z^2]^2}ds -\dfrac{2z^2}{\pi}\int_{-b}^{a}\dfrac{q(s)(x-s)}{[(x-s)^2 + z^2]^2}ds$}
\end{equation}

\begin{equation}
	\resizebox{\hsize}{!}{$\tau_{xz}^m = -\dfrac{2z^2}{\pi}\int_{-b}^{a}\dfrac{p(s)(x-s)}{[(x-s)^2 + z^2]^2}ds -\dfrac{2z}{\pi}\int_{-b}^{a}\dfrac{q(s)(x-s)^2}{[(x-s)^2 + z^2]^2}ds$}
\end{equation}

where the normal and tangential stress distributions, $p(s)$ and $q(s)$, have been assumed to be for a 2-D hertzian contact, with $q(s)$ proportional to $p(s)$, as proposed by McDowell \cite{mcdowell}:

\begin{equation}
	p(s) = p_0 \sqrt{1 - \left(s/a\right)^2}
\end{equation}

\begin{equation}
	q(s) = q_0 \sqrt{1 - \left(s/a\right)^2}
\end{equation}

where, $q_0 = -\mu p_0$, and $\mu$ is the coefficient of friction. The mechanical stress components obtained from these integrals are then transformed from the $(x', z')$ coordinate system of the shear zone, to the $(x, z)$ coordinate system of the workpiece, as proposed by \cite{su}.

\subsection{Thermal stresses}

The thermal stresses are derived using a model established by Liang et al. \cite{liang1, liang2}. The equations for calculating the thermal stress components $\sigma_{xx}^t$, $\sigma_{zz}^t$ and $\tau_{xz}^t$ are:

\begin{equation}
	\resizebox{\hsize}{!}{$
		\begin{split}
			\sigma_{xx}^t (x, z) &= -\dfrac{\alpha E}{1 - 2\nu}\int_{0}^{\infty}\int_{-\infty}^{\infty}{\bigl( G_{xh}(x',z')\dfrac{\partial T}{\partial x}(x,z) + G_{x\nu}(x',z')\dfrac{\partial T}{\partial z}(x,z)\bigr)dx' dz'}\\
			&+ \dfrac{2z}{\pi}\int_{-\infty}^{\infty}\dfrac{p(t)(t-x)^2}{((t-x)^2 + z^2)^2}dt - \dfrac{\alpha ET(x,z)}{1 - 2\nu}
		\end{split}
	$}
\end{equation}

\begin{equation}
	\resizebox{\hsize}{!}{$
		\begin{split}
			\sigma_{zz}^t (x, z) &= -\dfrac{\alpha E}{1 - 2\nu}\int_{0}^{\infty}\int_{-\infty}^{\infty}{\bigl( G_{zh}(x',z')\dfrac{\partial T}{\partial x}(x,z) + G_{z\nu}(x',z')\dfrac{\partial T}{\partial z}(x,z)\bigr)dx' dz'}\\
			&+ \dfrac{2z^3}{\pi}\int_{-\infty}^{\infty}\dfrac{p(t)}{((t-x)^2 + z^2)^2}dt - \dfrac{\alpha ET(x,z)}{1 - 2\nu}
		\end{split}
	$}
\end{equation}

\begin{equation}
	\resizebox{\hsize}{!}{$
		\begin{split}
			\tau_{xz}^t (x, z) &= -\dfrac{\alpha E}{1 - 2\nu}\int_{0}^{\infty}\int_{-\infty}^{\infty}{\bigl( G_{xzh}(x',z')\dfrac{\partial T}{\partial x}(x,z) + G_{xz\nu}(x',z')\dfrac{\partial T}{\partial z}(x,z)\bigr)dx' dz'}\\
			&+ \dfrac{2z^2}{\pi}\int_{-\infty}^{\infty}\dfrac{p(t)(t-x)}{((t-x)^2 + z^2)^2}dt
		\end{split}
	$}
\end{equation}

where $\alpha$ is the coefficient of thermal expansion of the workpiece, $E$ is the Young's modulus, $\nu$ is the poisson ratio, and $G_{xh}$, $G_{x\nu}$, $G_{zh}$, $G_{z\nu}$, $G_{xzh}$ and $G_{xz\nu}$ are Green's functions, calculated using their corresponding expressions developed by Saif et al. \cite{saif}. The function $p(t)$ represents surface traction, and is given by Liang et al. \cite{liang1, liang2} as:

\begin{equation}
	p(t) = \dfrac{\alpha ET(x=0, z=0)}{1 - 2\nu}
\end{equation}

Using the mechanical and thermal stresses obtained, the total stress components are calculated for elastic loading condition as:

\begin{equation}
	\begin{split}
		\sigma_{xx} &= \sigma_{xx}^m + \sigma_{xx}^t \\
		\sigma_{zz} &= \sigma_{zz}^m + \sigma_{zz}^t \\
		\tau_{xz} &= \tau_{xz}^m + \tau_{xz}^t \\
		\sigma_{yy} &= \nu (\sigma_{xx} + \sigma_{zz}) - \alpha E \Delta T
	\end{split}
\end{equation}

The model proposed by Shan et al. \cite{shan}, based on the Jiang-Sehitoglu \cite{sehitoglu} model, has been incorporated for calculating the stress and strain increments for elasto-plastic loading. The stress increments are obtained through an iterative process, which involves the use of deviatoric stresses, the shear yield strength, and the plastic modulus function (denoted by $h$).

\subsection{Stress relaxation to obtain residual stresses}

After the cutting process is complete, the unloading (relaxation) of the stresses must be performed in order to satisy certain boundary conditions for the stresses and strains. The stress and strain increments for this process are proposed by Merwin and Johnson \cite{merwinjohnson}, and Su \cite{su}:

\begin{equation}
	\begin{split}
		\Delta\sigma_{zz}^r &= -\dfrac{\sigma_{zz}^r}{M}, \Delta\tau_{xz}^r = -\dfrac{\tau_{xz}^r}{M} \\
		\Delta\varepsilon_{xx}^r &= -\dfrac{\varepsilon_{xx}^r}{M}, \Delta T^r = -\dfrac{T^r}{M}
	\end{split}
\end{equation}

Using these increments, either elastic or elastic-plastic unloading is performed, based on the work of Shan et al. \cite{shan}, which has also been proposed by Su \cite{su}, McDowell \cite{mcdowell}, and Liang et al. \cite{liang1}.

\section{Validation}

\renewcommand{\arraystretch}{1.2} 
\begin{table}[b]
	\centering
	\caption{Process parameters taken for the model}
	\label{tab:1}
	\begin{tabular*}{\linewidth}{l @{\extracolsep{\fill}} l}
		\toprule
		Parameters                             & Values          \\ \midrule
		Tool rake angle, $\alpha$              & $12.62^{\circ}$ \\
		Width of cut, $w$                      & 2 mm            \\
		Nose radius, $r_e$                     & $30\, \mu m$      \\
		Flank wear length, VB                  & $10\, \mu m$      \\
		$\psi$ (in oxley temperature model)          & 0.4             \\
		$\eta$ (in oxley temperature model)          & 0.9             \\
		Friction factor, $m_{plow}$            & 0.99            \\
		$\rho_{plow}$ (in plowing force model) & $20^{\circ}$   \\ \botrule
	\end{tabular*}
\end{table}

\begin{table*}[]
	\centering
	\caption{Properties of Inconel 718 and Ti6Al4V for material parameters}
	\label{tab:2}
	\begin{tabular*}{\linewidth}{l @{\extracolsep{\fill}} ll}
		\toprule
		Properties                                          & Inconel 718           & Ti6Al4V              \\ \midrule
		Melting temperature, $T_m$                          & $1297^{\circ}C$       & $1604^{\circ}C$      \\
		Density, $\rho$ (in $kg/m^3$)                       & 8280                  & 4430                 \\
		Specific heat capacity, $S$ (in $J / kg K$)         & 362                   & 565                  \\
		Thermal conductivity, $k$ (in $W/mK$)               & 10.3                  & 6.6                  \\
		Thermal diffusivity, $a$ (in $m^2 / s$)             & $2.87\times 10^{-6}$  & $2.76\times 10^{-6}$ \\
		Young's modulus, $E$                                & 214.580 GPa           & 113.8 GPa            \\
		Poisson ratio, $\nu$                                & 0.305                 & 0.342                \\
		Coefficient of friction, $\mu$                      & 0.7                   & 0.85                 \\
		Shear yield stress                                  & $1100 / \sqrt{3}$ MPa & $880 / \sqrt{3}$ MPa \\
		Coefficient of thermal expansion, $\alpha$ (in $K^{-1}$) & $14.8\times 10^{-6}$  & $9.2\times 10^{-6}$ \\ \botrule
	\end{tabular*}
\end{table*}

The authors have utilized machining data associated with two work materials - Inconel 718 and Ti6Al4V - to validate the proposed residual stress model. These materials have been selected due to their widespread application in the aerospace industry, particularly in critical components such as turbine blades and various structural parts, owing to their exceptional mechanical properties, thermal stability, and corrosion resistance. The parameters for the Johnson-Cook model adopted for Inconel 718 are as follows: $A = 1290\, \text{MPa}$, $B = 895\, \text{MPa}$, $C = 0.0252$, $n = 0.526$, $m = 1.55$, and $\dot{\varepsilon}_0 = 10^{-5}$; for Ti6Al4V, they are: $A = 724.7\, \text{MPa}$, $B = 683.1\, \text{MPa}$, $C = 0.01$, $n = 0.47$, $m = 1.0$, and $\dot{\varepsilon}_0 = 10^{-5}$. The ambient temperature $T_0$ is taken as $20^{\circ} C$ (equivalent to 293 K). The parameters for the Johnson-Cook model for Inconel 718 are derived from the work of Zhou and Ren \cite{zhou-in}, while those for Ti6Al4V are obtained from Zhou et al. \cite{zhou-ti}. The process parameters utilized for the model are displayed in Table \ref{tab:1}, and the material properties of Inconel 718 and Ti6Al4V are presented in Table \ref{tab:2}.

The values for the tool rake angle, width of cut, nose radius, density, specific heat, thermal conductivity, coefficient of friction, $m_{plow}$, $\rho_{plow}$, melting temperature, $\psi$, and $\eta$ for both Inconel 718 and Ti6Al4V are sourced from the data provided by Zhou and Ren \cite{zhou-in} and Zhou et al. \cite{zhou-ti}. The values for the Young's modulus, Poisson ratio, and coefficient of thermal expansion for Inconel 718 are obtained from Shen et al. \cite{shen}, while those for Ti6Al4V are derived from Shan et al. \cite{shan}. The thermal diffusivity of Ti6Al4V is sourced from Liang et al. \cite{liang2}.

The tool is modeled as an uncoated cemented carbide tool, with a density of $14860\, kg/m^3$, thermal conductivity of $82\, W/mK$, and specific heat of $249.8\, J / kg K$, based on data provided by Liu et al. \cite{liu}. The coolant is modeled as water, with an absolute viscosity of $0.001\, kg/m.s$, Prandtl number of 7.56, and a diameter of the circular duct of 2 mm, as per data from Singh and Sharma \cite{singh}. For this model, a mass flow rate per unit area of $2500\, kg / m^2 s$ is considered to model the effect of the coolant on the temperature distribution.

In the force model, the shear angle $\phi$ is iteratively computed within the range of $10^{\circ}$ to $45^{\circ}$. The parameters $C_0$ and $\delta$ are treated as hyperparameters, which are optimized by tuning them using the training data. Specifically, $C_o$ is varied within the range of 0.1 to 4.0, while $\delta$ is varied within the range of 0.005 to 0.20. The cost function employed for determining the optimal values of these hyperparameters is $(1+C_{mpe})(1+T_{mpe})(1+C_{sde})(1+T_{sde})$, where $C_{mpe}$ and $T_{mpe}$ denote the mean absolute percentage errors in the cutting and thrust forces, respectively, and $C_{sde}$ and $T_{sde}$ represent the standard deviation of errors in the cutting and thrust forces. The experimental data utilized for training the model, and subsequently evaluating it to compute the errors in $F_c$ and $F_t$, are obtained from Zhou and Ren \cite{zhou-in} (for Inconel 718) and Zhou et al. \cite{zhou-ti} (for Ti6Al4V). The plastic modulus function $h$ is tuned within the range of 10 GPa to 60 GPa, employing a mean absolute percentage error cost function to minimize the error in the predicted residual stress. For stress relaxation, the number of iterations $M$ is set to 100.

The integrals employed in temperature and stress modeling are computed using the quad and dblquad functions in the Python scipy-1.7.3 library. The model is implemented in Python-3.9.12 and executed on a 64-bit system featuring an Intel(R) Core(TM) i7-8565U processor, with a clock speed of 1.80 GHz and 16 GB RAM.

\section{Results and discussion}

\begin{figure}
	\includegraphics[width=0.5\textwidth]{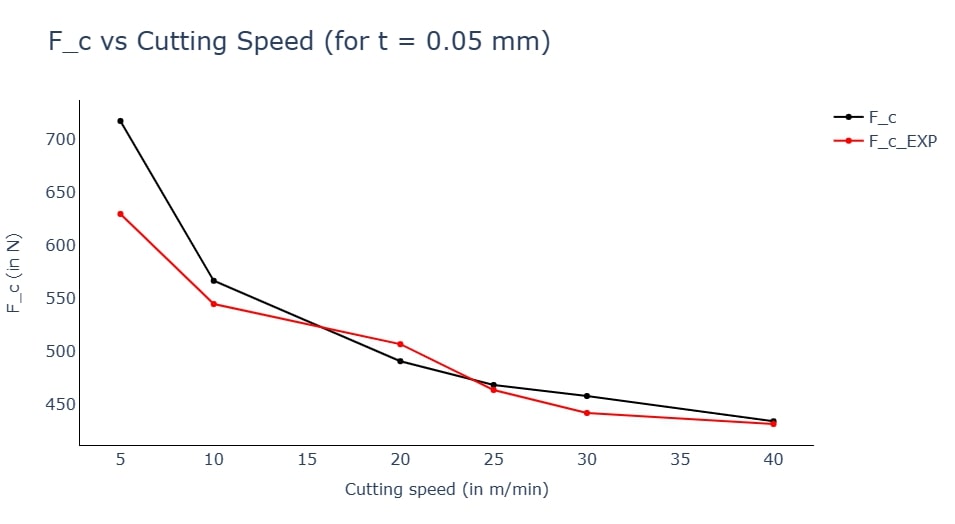}\hfill
	\includegraphics[width=0.5\textwidth]{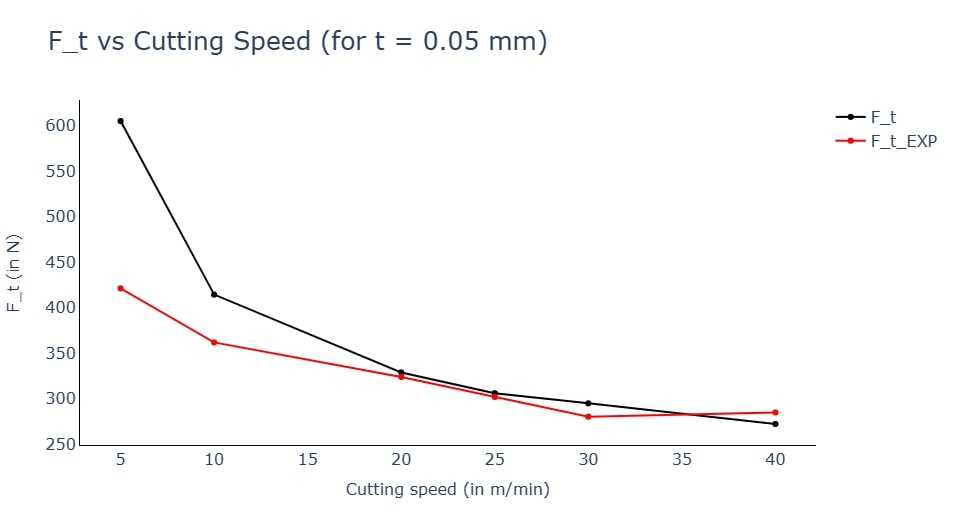}
	\includegraphics[width=0.5\textwidth]{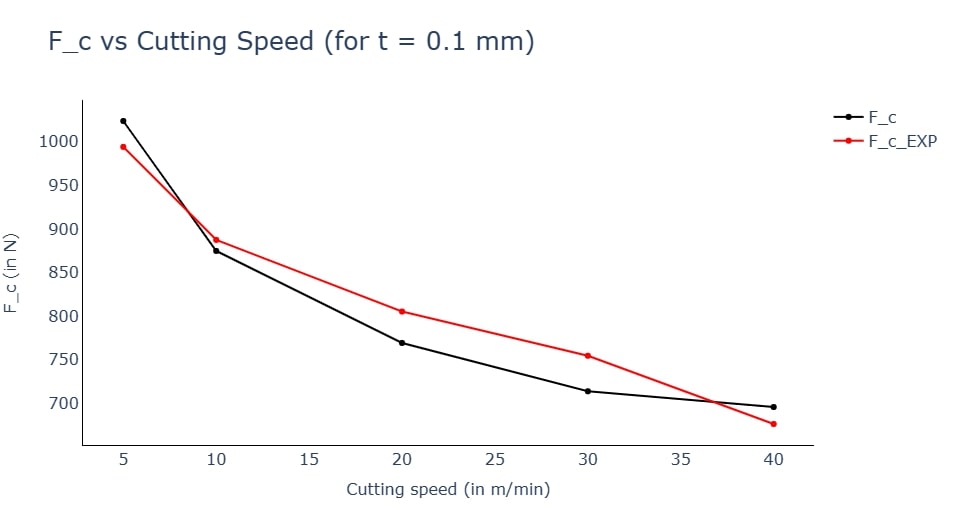}\hfill
	\includegraphics[width=0.5\textwidth]{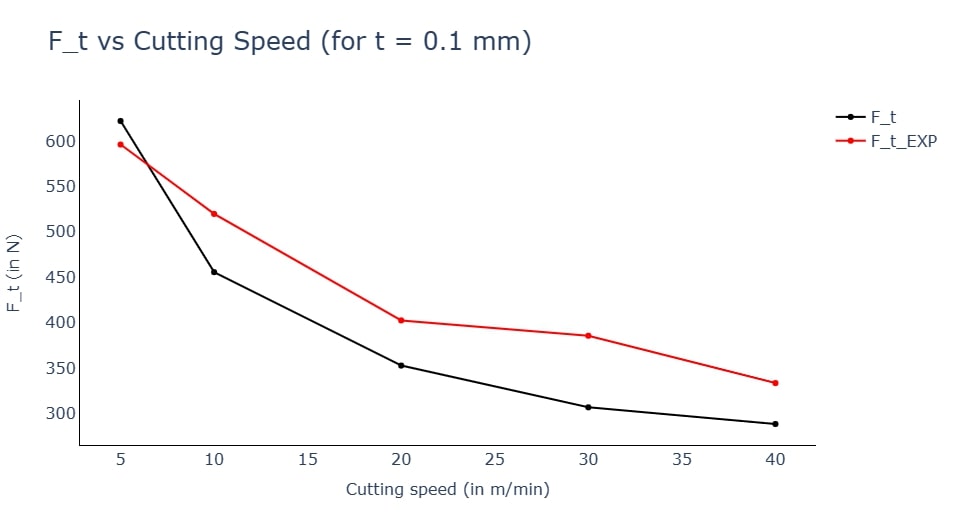}
	\caption{Force vs cutting speed plots at t = 0.05 mm and t = 0.10 mm for Inconel 718. The black lines represent model predictions, and the red lines are from experimental data.}
	\label{fig:2}
\end{figure}

\begin{figure}
	\includegraphics[width=0.5\textwidth]{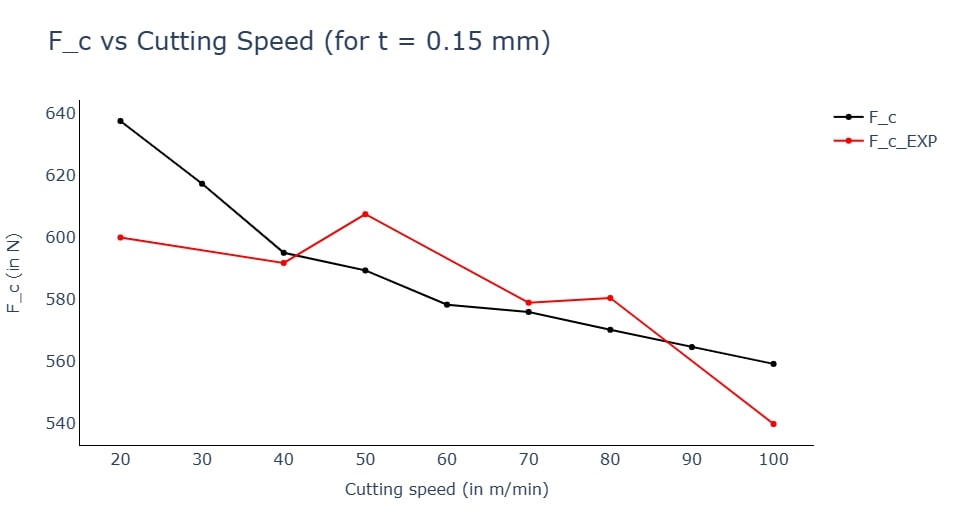}\hfill
	\includegraphics[width=0.5\textwidth]{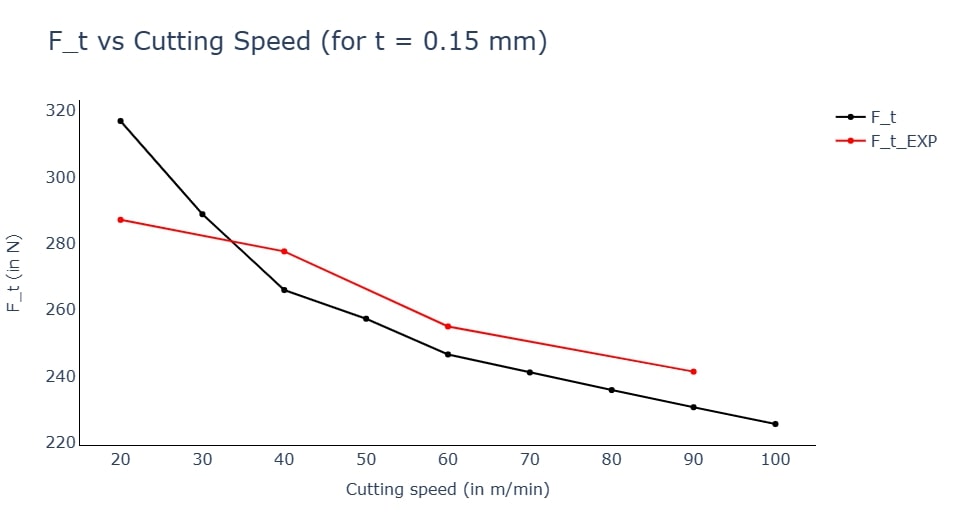}
	\caption{Force vs cutting speed plots at t = 0.15 mm for Ti6Al4V. The black lines represent model predictions, and the red lines are from experimental data.}
	\label{fig:3}
\end{figure}

\begin{figure}
	\includegraphics[width=0.5\textwidth]{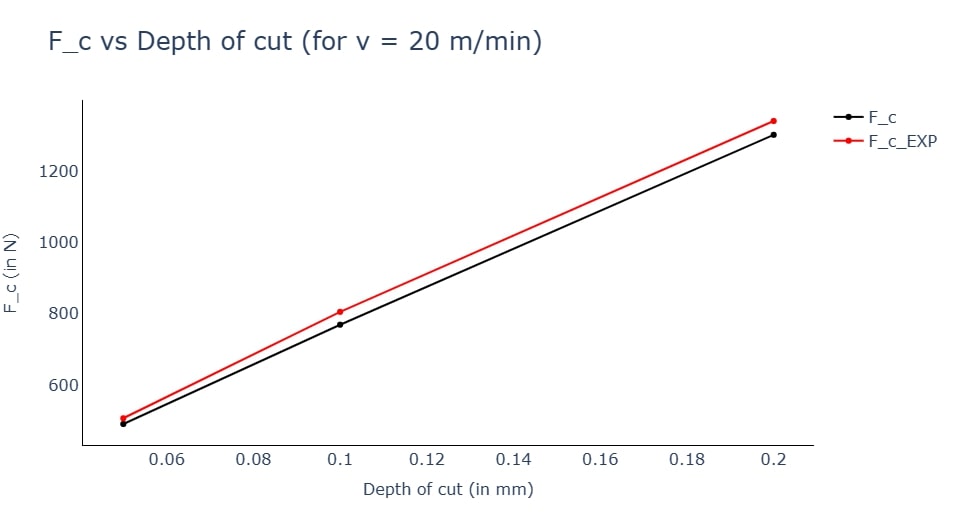}\hfill
	\includegraphics[width=0.5\textwidth]{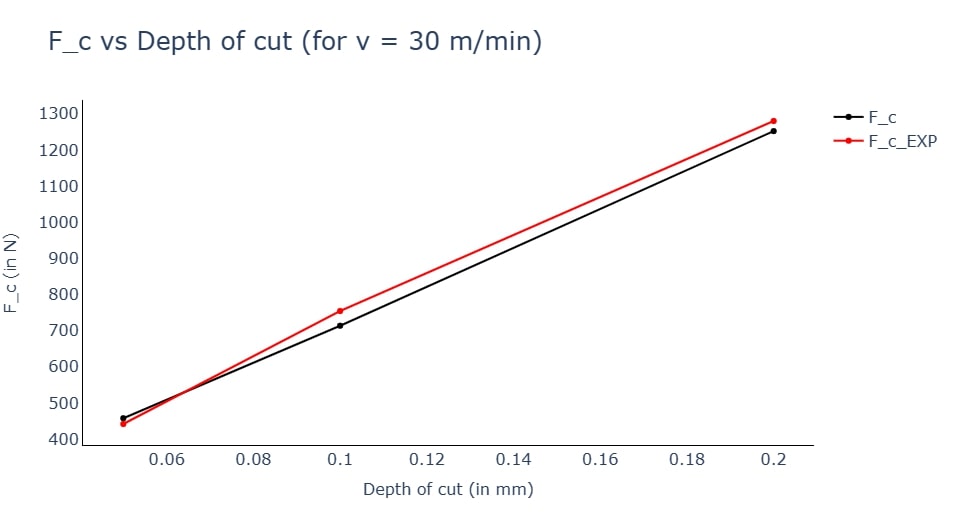}
	\caption{Cutting force vs depth of cut plots at v = 20 m/min and v = 30 m/min for Inconel 718. The black lines represent model predictions, and the red lines are from experimental data.}
	\label{fig:4}
\end{figure}

\begin{figure}
	\includegraphics[width=0.5\textwidth]{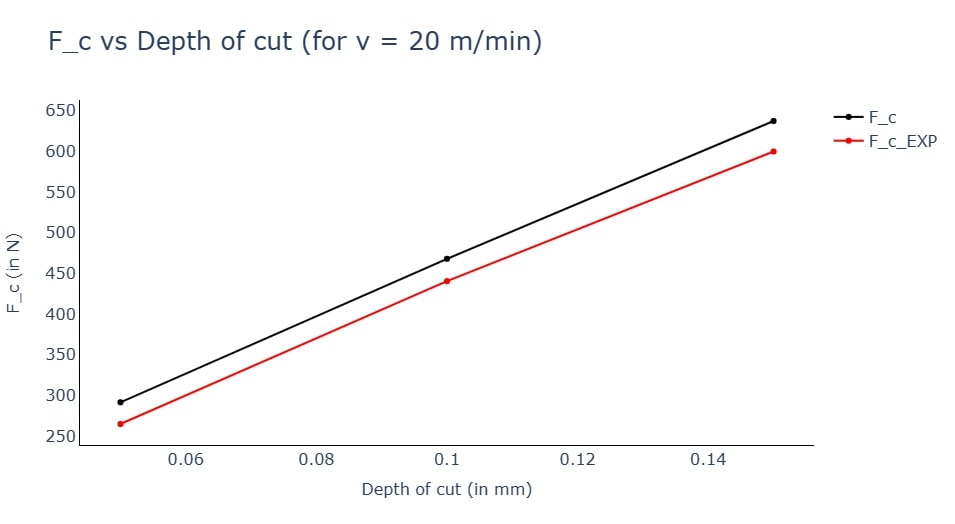}\hfill
	\includegraphics[width=0.5\textwidth]{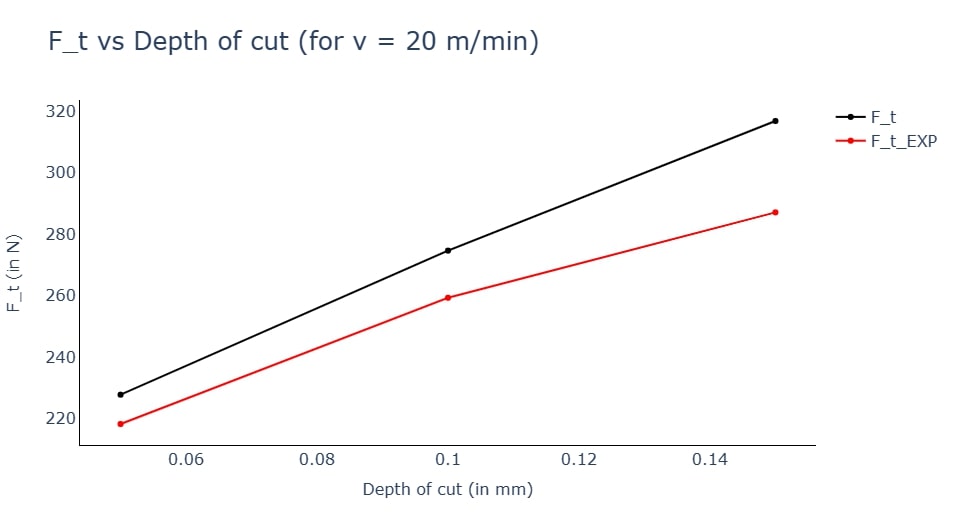}
	\caption{Force vs depth of cut plots at v = 20 m/min for Ti6Al4V. The black lines represent model predictions, and the red lines are from experimental data.}
	\label{fig:5}
\end{figure}

\begin{figure}
	\includegraphics[width=0.5\textwidth]{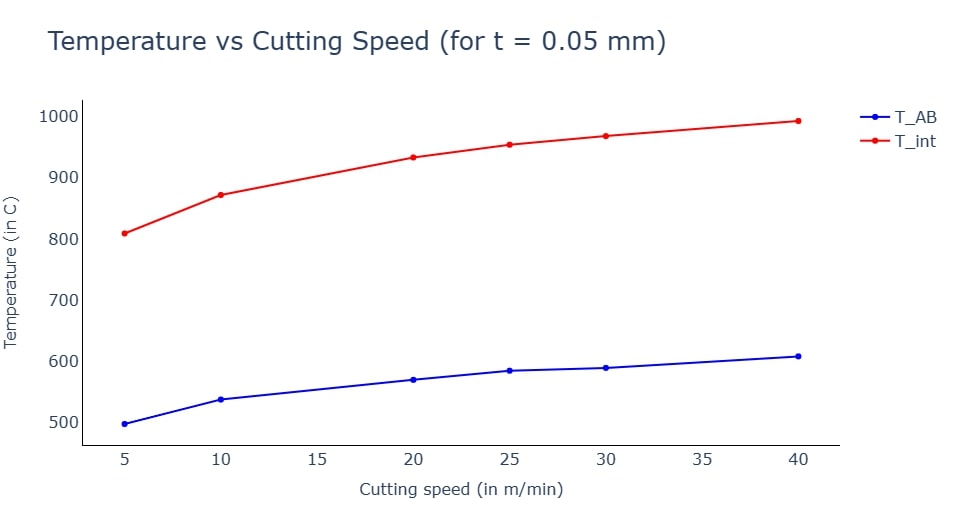}\hfill
	\includegraphics[width=0.5\textwidth]{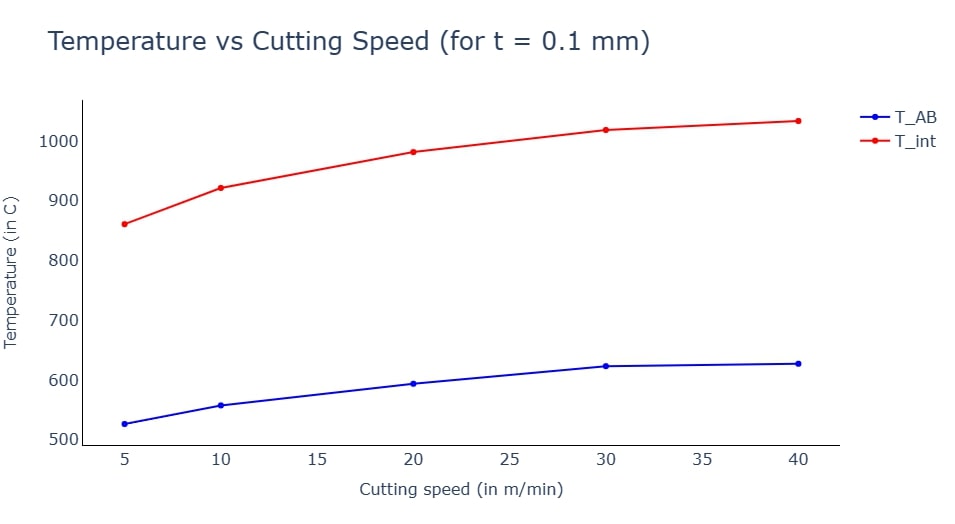}
	\includegraphics[width=0.5\textwidth]{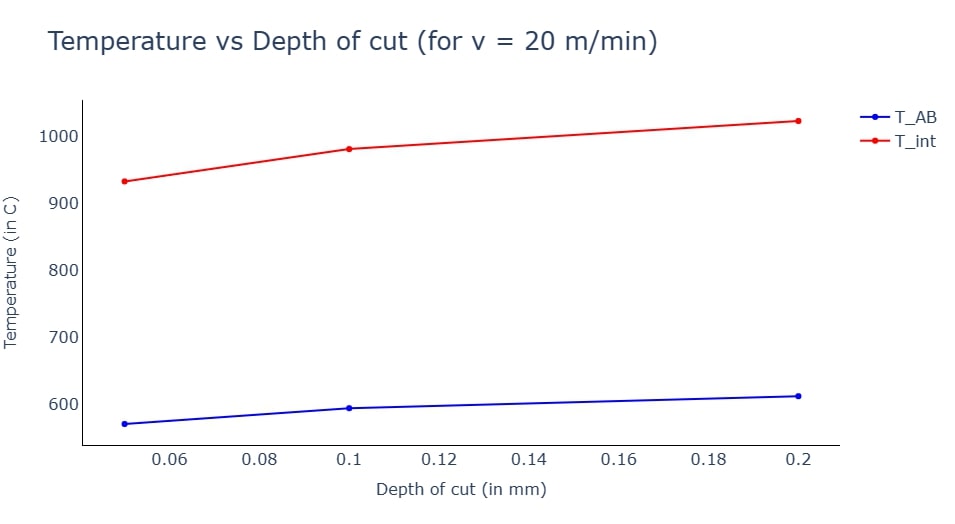}\hfill
	\includegraphics[width=0.5\textwidth]{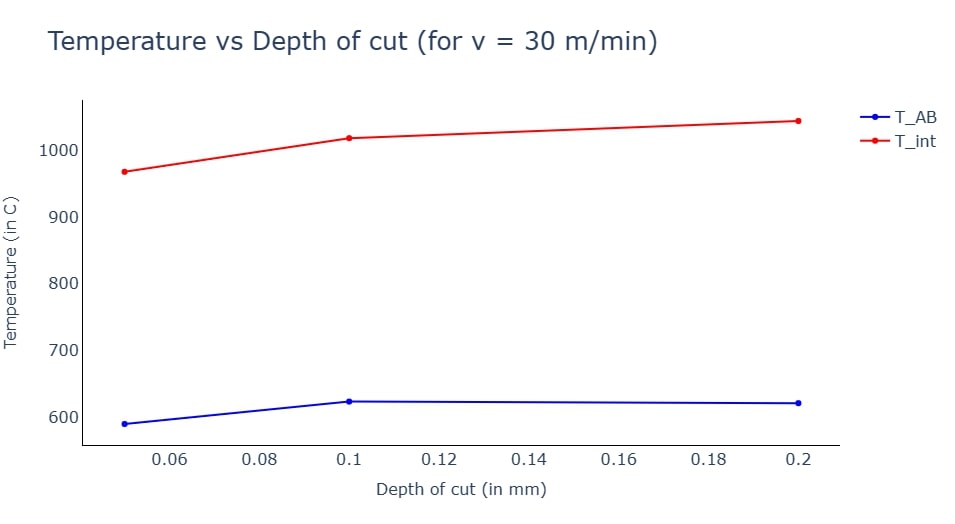}
	\caption{Temperature vs cutting speed, and temperature vs depth of cut plots for Inconel 718. The blue lines represent predicted shear zone temperature ($T_{AB}$), and the red lines represent predicted tool-chip interface temperature ($T_{int}$).}
	\label{fig:6}
\end{figure}

\begin{figure}
	\includegraphics[width=0.5\textwidth]{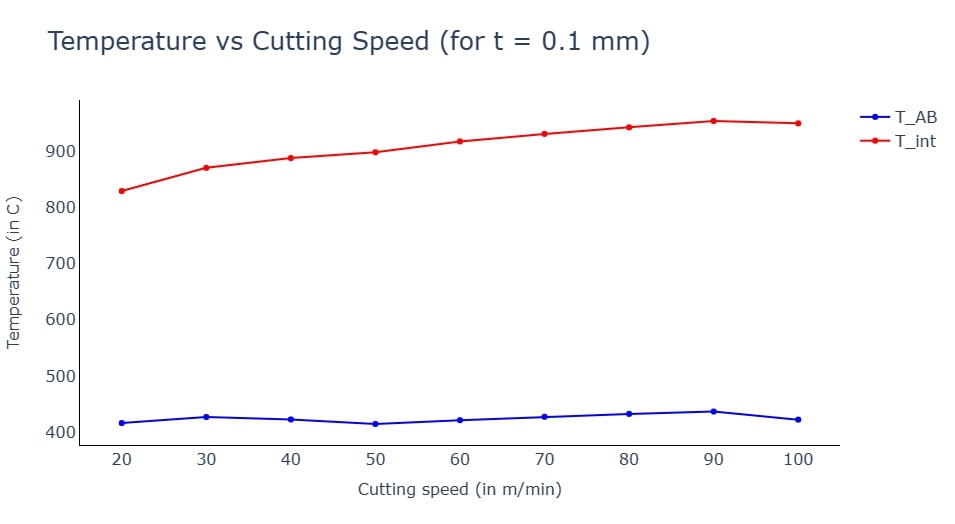}\hfill
	\includegraphics[width=0.5\textwidth]{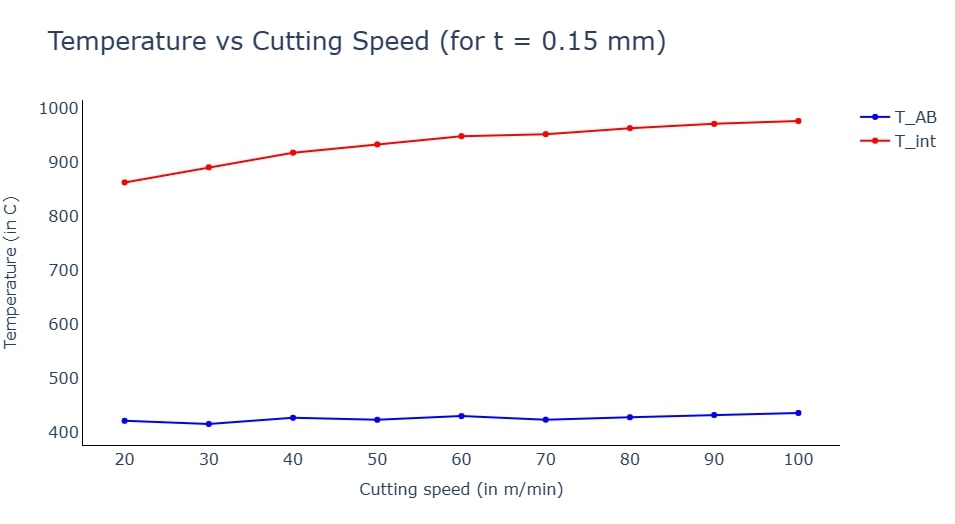}
	\includegraphics[width=0.5\textwidth]{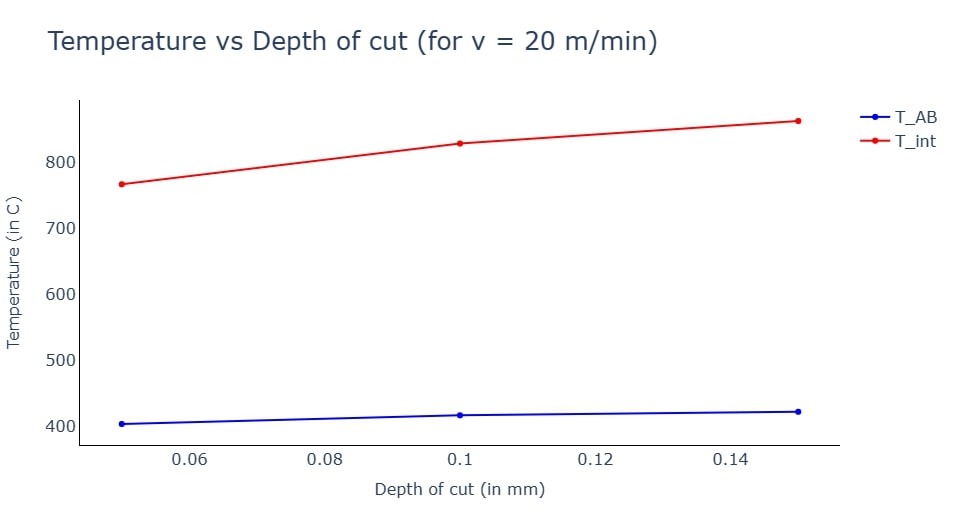}\hfill
	\includegraphics[width=0.5\textwidth]{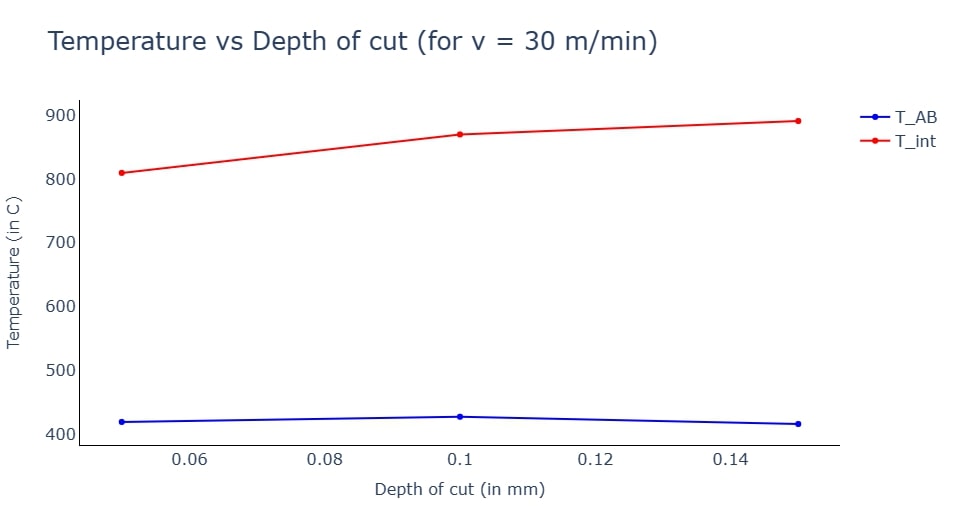}
	\caption{Temperature vs cutting speed, and temperature vs depth of cut plots for Ti6Al4V. The blue lines represent predicted shear zone temperature ($T_{AB}$), and the red lines represent predicted tool-chip interface temperature ($T_{int}$).}
	\label{fig:7}
\end{figure}

\begin{figure}
	\includegraphics[width=0.5\textwidth]{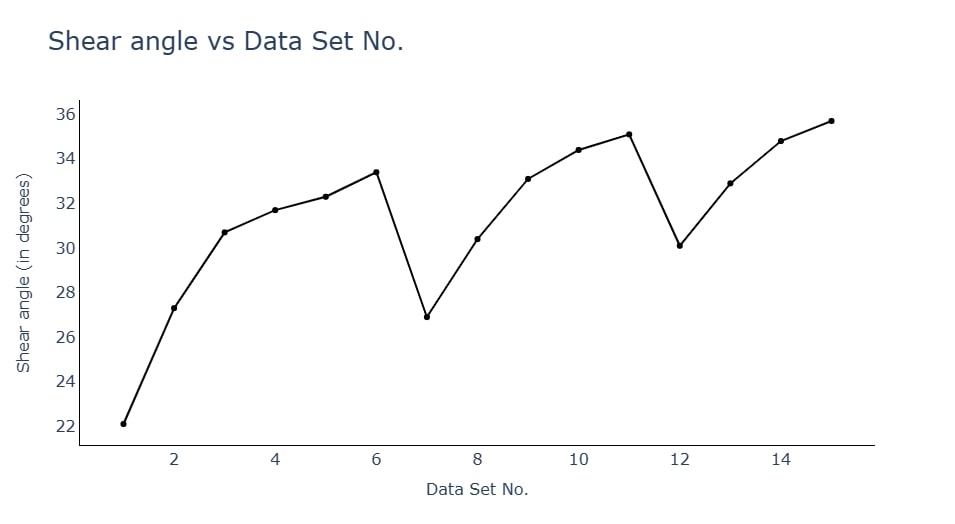}\hfill
	\includegraphics[width=0.5\textwidth]{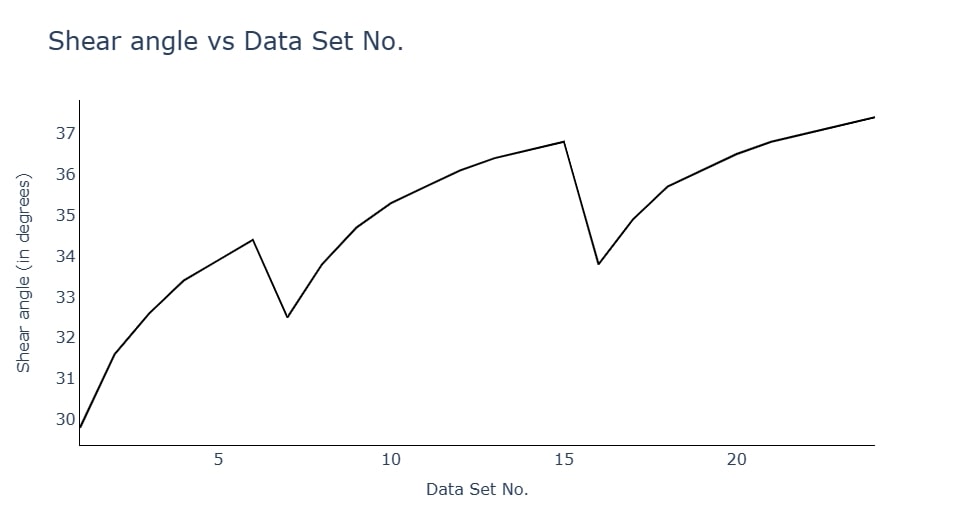}
	\caption{Shear angles predicted for each v-t pair for Inconel 718 \cite{zhou-in} (top), and Ti6Al4V \cite{zhou-ti} (bottom).}
	\label{fig:8}
\end{figure}

\begin{figure}
	\includegraphics[width=0.5\textwidth]{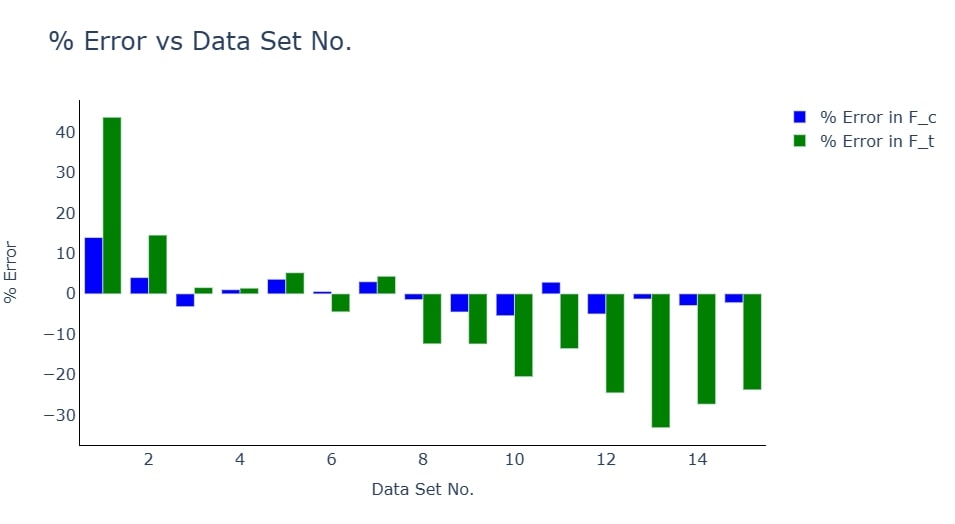}\hfill
	\includegraphics[width=0.5\textwidth]{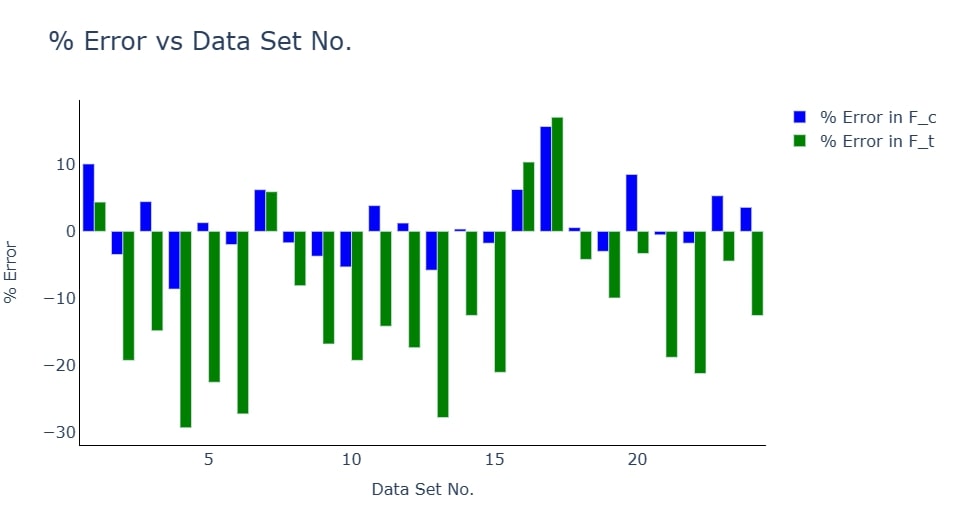}
	\caption{Errors in $F_c$ and $F_t$ calculated for each v-t pair for Inconel 718 \cite{zhou-in} (top), and Ti6Al4V \cite{zhou-ti} (bottom). The blue bars represent the percentage error in $F_c$, and green bars represent percentage error in $F_t$.}
	\label{fig:9}
\end{figure}

\begin{figure}
	\includegraphics[width=0.5\textwidth]{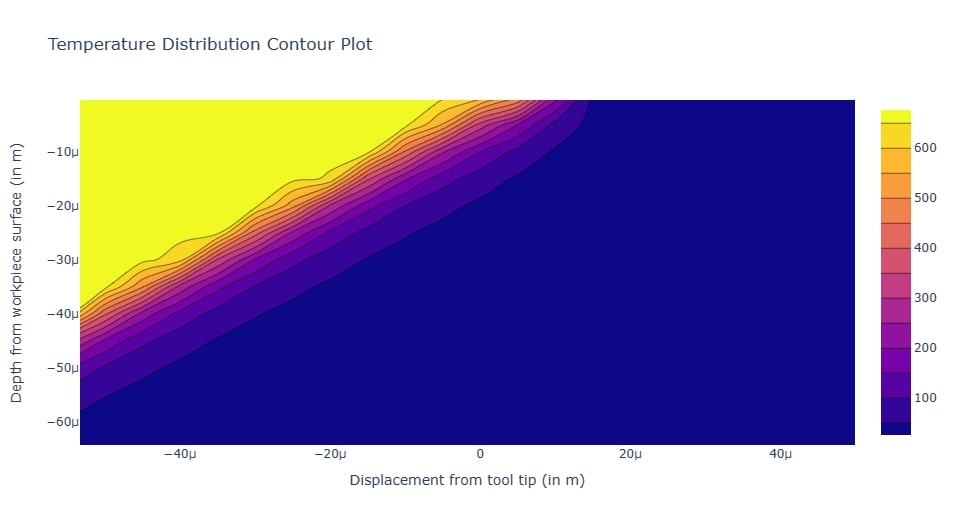}\hfill
	\includegraphics[width=0.5\textwidth]{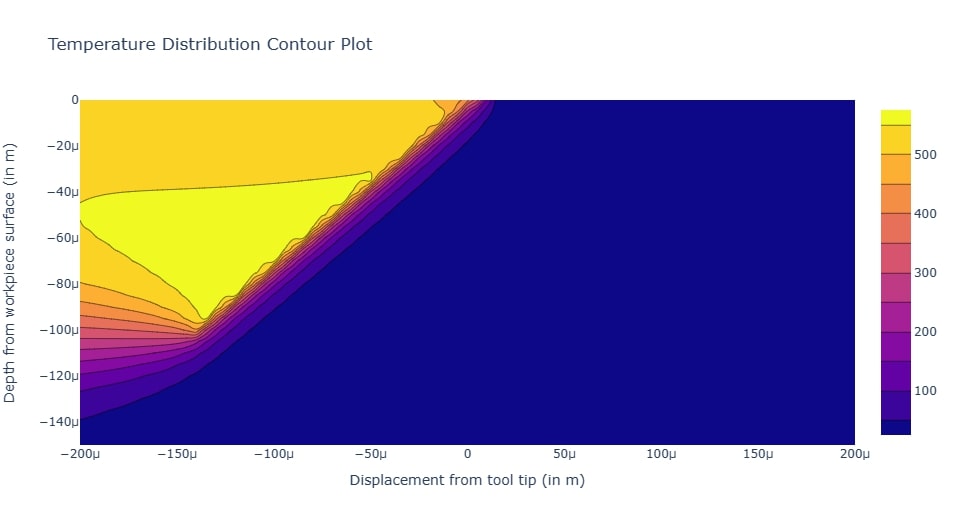}
	\caption{The predicted temperature distribution through the workpiece, for Inconel 718 (top) and Ti6Al4V (bottom).}
	\label{fig:10}
\end{figure}

\begin{figure}
	\includegraphics[width=0.5\textwidth]{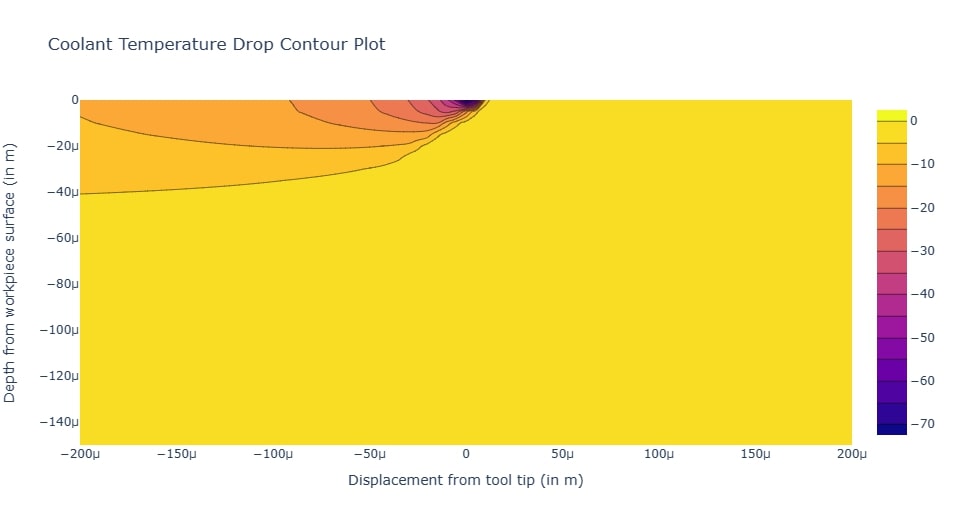}\hfill
	\includegraphics[width=0.5\textwidth]{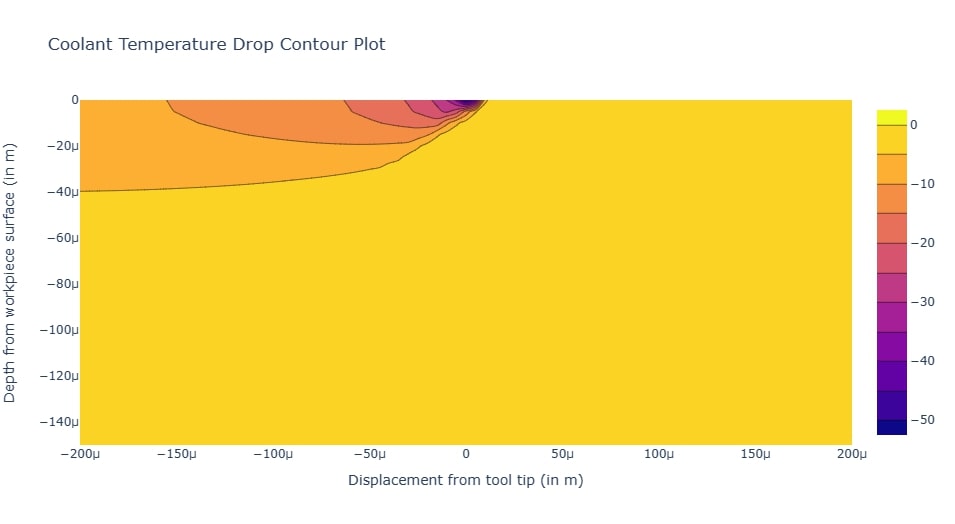}
	\caption{The predicted change in temperature through the workpiece due to the coolant, for Inconel 718 (top) and Ti6Al4V (bottom).}
	\label{fig:11}
\end{figure}

\begin{figure}
	\includegraphics[width=0.5\textwidth]{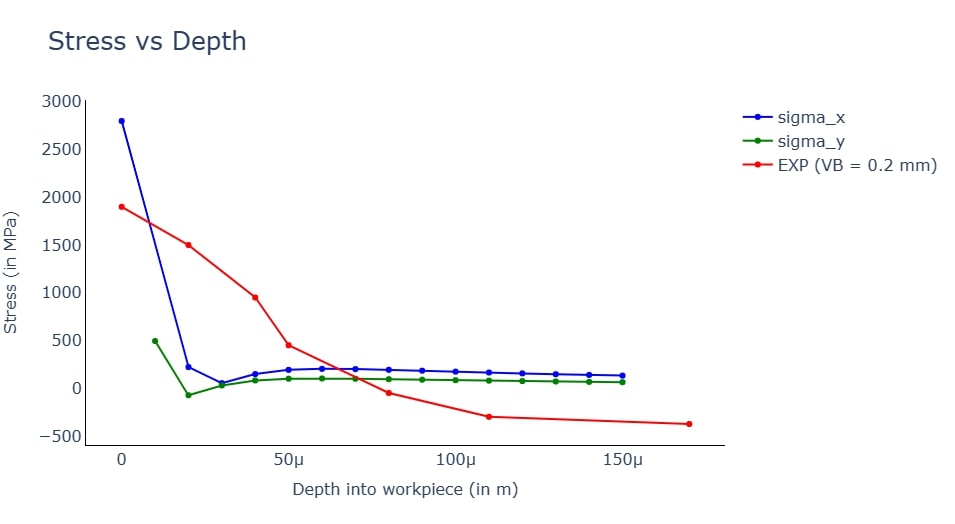}\hfill
	\includegraphics[width=0.5\textwidth]{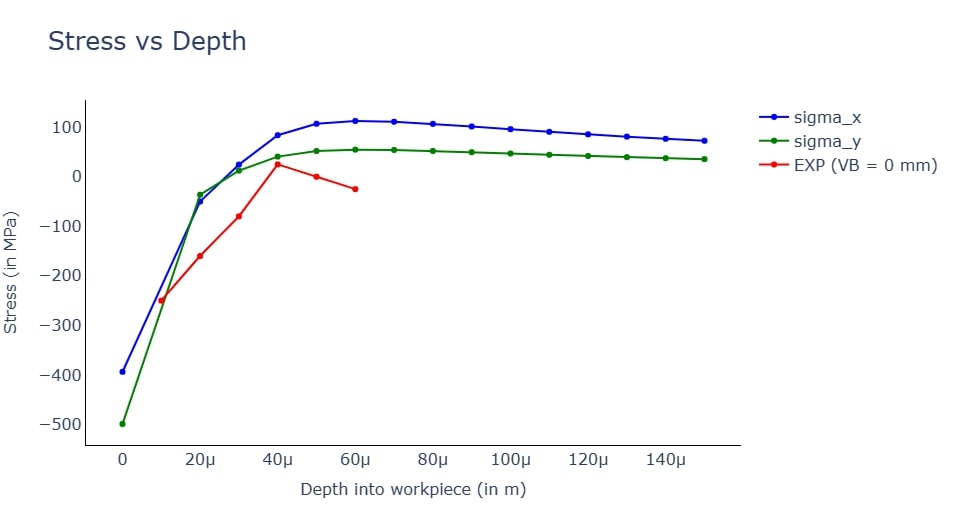}
	\caption{The predicted residual stress vs depth into the workpiece, for Inconel 718 (top) and Ti6Al4V (bottom). The red line represents experimental data, and the blue and green lines represent $\sigma_{xx}$ and $\sigma_{yy}$ respectively.}
	\label{fig:12}
\end{figure}

\begin{figure}
	\includegraphics[width=0.5\textwidth]{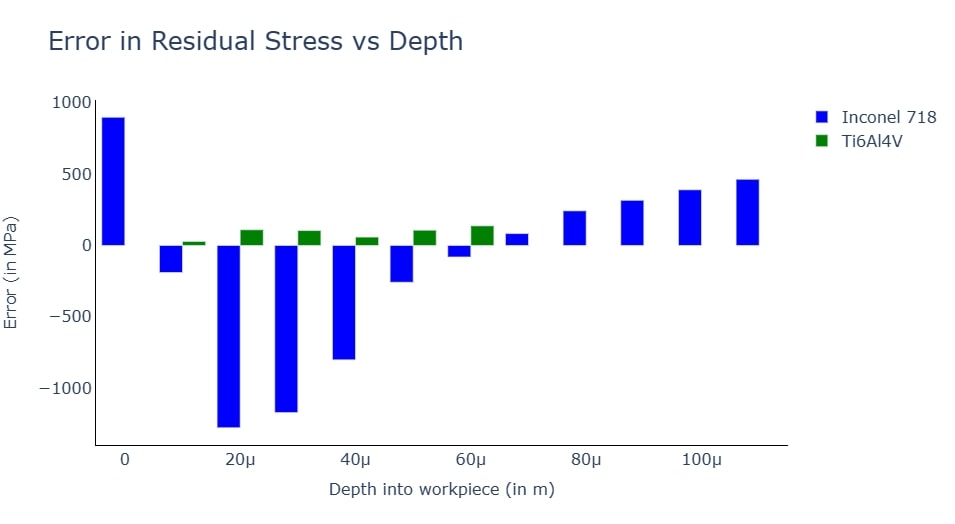}
	\caption{Errors in residual stress vs depth into the workpiece, for Inconel 718 (blue bars) and Ti6Al4V (green bars).}
	\label{fig:13}
\end{figure}

The force model is employed for both Inconel 718 and Ti6Al4V, utilizing the parameters detailed in Tables \ref{tab:1}-\ref{tab:2}. For Inconel 718, the predicted cutting forces ($F_c$) and thrust forces ($F_t$) yield mean percentage errors of 3.65\% and 16.16\%, respectively, with standard deviations of the errors amounting to 24.46 N and 79.31 N, respectively. Correspondingly, for Ti6Al4V, the errors stand at 4.37\% for $F_c$ and 15.12\% for $F_t$, with standard deviations of 17.65 N and 20.05 N, respectively. Generally, the force model demonstrates higher accuracy in predicting cutting force compared to thrust force, a trend in line with observations by Zhou and Ren \cite{zhou-in} regarding the superior accuracy of cutting force prediction with various J-C models. As proposed by Zhou and Ren \cite{zhou-in}, the predictive accuracy may be enhanced by employing an improved plow force model.

Plots illustrating cutting force and thrust force against speed ($v$) and depth of cut ($t$) are provided in Figures \ref{fig:2}-\ref{fig:5} for both Inconel 718 and Ti6Al4V. Figures \ref{fig:6} and \ref{fig:7} depict plots for the average shear zone temperatures ($T_{AB}$) and average tool-chip interface temperatures ($T_{int}$) computed from Boothroyd's \cite{boothroyd} model, in relation to cutting speed and depth of cut for both materials. Furthermore, Figure \ref{fig:8} illustrates the predicted shear angles obtained from the force model for Inconel 718 and Ti6Al4V, across all data sets (numbered 1-15 in \cite{zhou-in} for Inconel 718 and 1-24 in \cite{zhou-ti} for Ti6Al4V). The percentage errors in $F_c$ and $F_t$ for each individual data set are plotted in Figure \ref{fig:9}. The values of $C_o$ and $\delta$ derived from the force model are 1.8 and 0.185 for Inconel 718, and 0.1 and 0.145 for Ti6Al4V.

Temperature distribution plots within the workpiece and the impact of coolant are presented in Figures \ref{fig:10} and \ref{fig:11} for both Inconel 718 and Ti6Al4V. The maximum temperature drop due to coolant, as observed in Figure \ref{fig:11}, falls within the range of $50-70^{\circ} C$, notably smaller than the anticipated drop (approximately 29\% for laminar and 53\% for turbulent flow, as suggested by \cite{singh}). Possible explanations for this disparity include: (a) The convective heat flow rate ($q_{cool}$) used in temperature drop calculations is directly determined for the workpiece, without consideration of the flow rate for the tool, a significant heat source. (b) The impact of coolant has not been incorporated into the force model, resulting in higher forces and consequently greater temperature values than expected in the presence of coolant.

Residual stresses ($\sigma_{xx}$ and $\sigma_{yy}$) are plotted alongside experimental data in Figure \ref{fig:12}. For Inconel 718, the results are plotted for a cutting speed of 60 m/min and a depth of cut of 0.15 mm, along with experimental data from Liu et al. \cite{liu}. Similarly, for Ti6Al4V, the results are plotted for a cutting speed of 50 m/min and a depth of cut of 0.10 mm, alongside experimental data from Liang et al. \cite{liang1}. The values of the plastic modulus function $h$ derived from the residual stress model are 40 GPa for Inconel 718 and 25 GPa for Ti6Al4V. The model results indicate that residual stresses are more pronounced near the surface, diminishing rapidly with increasing depth, consistent with the expectation that mechanical and thermal effects of machining primarily influence regions near the surface. Tensile stress regions are observed near the surface for Inconel 718, attributed to the dominance of tensile thermal effects over compressive mechanical effects as explained by Liu et al. \cite{liu}. This observation aligns with the maximum temperatures for Inconel 718 ($> 600^{\circ}$C), higher than those for Ti6Al4V ($> 500^{\circ}$C), partly due to larger $v$ and $t$ values used for modeling residual stresses in Inconel 718. Conversely, compressive stress regions near the surface are observed for Ti6Al4V, attributed to the dominance of mechanical effects over thermal effects, as explained by Liang et al. \cite{liang1}, given the lower heating due to friction for low cutting speeds (here, $v = 50$ m/min) and smaller flank wear (here, $\text{VB}_{model} = 10\,\mu m$, $\text{VB}_{exp} = 0\, mm$). The errors in residual stress (i.e., the difference between model prediction and experimental data) are plotted against depth into the workpiece for both Inconel 718 and Ti6Al4V in Figure \ref{fig:13}. The residual stress results exhibit similar trends as the experimental data but show larger errors compared to the force model. This discrepancy can partly be attributed to the fact that the parameters (aside from $v$ and $t$) for which the experimental data are plotted differ from those used in modeling residual stresses. Notably, variations in experimental values with parameters such as flank wear (VB) are observed in results provided by Liu et al. \cite{liu} and Liang et al. \cite{liang1}. Additionally, the accumulation of errors from component force, temperature, and stress analytical models contributes to larger errors.

\section{Conclusions}

This study introduces a physics-informed and data-driven model designed to predict residual stresses arising from orthogonal machining. Employing analytical modeling techniques, the authors model the forces and temperature distributions, facilitating the subsequent computation of residual stress variations with depth into the workpiece. Upon comparison with experimental data, the force model demonstrates precise predictions for $F_c$ and $F_t$ values. Although the residual stress exhibits larger errors, it aligns with anticipated trends from literature for both materials, considering their respective input parameters - notably, a tensile region in Inconel 718 and a compressive region in Ti6Al4V, closer to the surface. Key factors integrated into the model encompass material properties, machining process parameters (including tool rake angle, width of cut, nose radius, and flank wear), and the influence of coolant on the workpiece owing to laminar and turbulent flow.

The proposed model exhibits adaptability for specific tool-work-process combinations even with limited data availability, enhancing its practical utility for industrial applications. This adaptability enables the prediction of parameters for which data is scarce across various materials. Moreover, the model holds promise for extension into a real-time framework for industrial deployment. Such real-time integration would facilitate continual verification of experimental measurements against their expected counterparts, facilitating adjustments for any deviations attributable to improper processing conditions, thereby enhancing machining efficiency and component quality. Future enhancements to the model could involve the inclusion of additional factors such as machining vibrations, the impact of coolant on both the force model and temperature profiles in the shear zone and tool-chip interface, utilization of refined literature data, and adoption of advanced optimization techniques for hyperparameter tuning, with the aim of improving predictive accuracy.

\backmatter

\bmhead{Acknowledgments}

The authors would like to thank TCS Research for supporting this work.

\section*{Declarations}

\bmhead{Author contributions}

Dhar, R.: methodology, investigation, model development, programming, formal analysis, resources, writing - original draft and editing. Krishna, A. and Muhammed, B.: study conception, methodology, resources, reviewing, supervision and project administration. All authors have extensively revised and approved the final manuscript.

\bmhead{Ethics approval}

The authors declare that this manuscript was not submitted to more than one journal for simultaneous consideration. Also, the submitted work was original and has not been published elsewhere in any form or language.

\bmhead{Competing interests}

The authors declare no competing interests.

\balance
\bibliography{sn-bibliography}

\end{document}